\documentclass{aa}  

\usepackage{graphicx}
\usepackage{txfonts}
\usepackage[switch]{lineno}
\bibpunct{(}{)}{;}{a}{}{,}
\usepackage{float}
\usepackage{hyperref}
\hypersetup{
    colorlinks=true,
    linkcolor=blue,
    filecolor=magenta,      
    urlcolor=blue,
    citecolor=blue,
}
\begin{document}

   \title{Hints of $\gamma$-ray orbital variability from $\gamma ^2$ Velorum}

   \author{G. Martí-Devesa
          \inst{1}
          \and
          O. Reimer\inst{1}
          \and
          J. Li\inst{2}
          \and
          D.F. Torres\inst{3,4,5}
          }

   \institute{Institut f\"{u}r Astro- und Teilchenphysik, Leopold-Franzens-Universit\"{a}t Innsbruck, A-6020 Innsbruck, Austria\\\email{guillem.marti-devesa@uibk.ac.at}
                \and
                        Deutsches Elektronen Synchrotron DESY, D-15738 Zeuthen, Germany
                \and
                        Institute of Space Sciences (ICE, CSIC), Campus UAB, Carrer de Magrans s/n, 08193 Barcelona, Spain
                \and
                        Instituci\'{o} Catalana de Recerca i Estudis Avan\c{c}ats (ICREA), E-08010 Barcelona, Spain
                \and
                        Institut d’Estudis Espacials de Catalunya (IEEC), 08034 Barcelona, Spain\\}

   \date{Received -, -; accepted -, -}

  \abstract
   {Colliding wind binaries are massive systems featuring strong, interacting stellar winds which may act as particle accelerators. Therefore, such binaries are good candidates for detection at high energies. However, only the massive binary $\eta$ Carinae has been firmly associated with a $\gamma$-ray signal. A second system, $\gamma^2$ Velorum, is positionally coincident with a $\gamma$-ray source, but we lack unambiguous identification.}
   {Observing orbital modulation of the flux would establish an unambiguous identification of the binary $\gamma ^2$ Velorum as the $\gamma$-ray source detected by the \textit{Fermi} Large Area Telescope (\textit{Fermi}-LAT).}
   {We used more than ten years of observations with \textit{Fermi}-LAT. Events are phase-folded with the orbital period of the binary to search for variability. We studied systematic errors that might arise from the strong emission of the nearby Vela pulsar with a more conservative pulse-gated analysis.}
   {We find hints of orbital variability, indicating maximum flux from the binary during apastron passage.}
   {Our analysis strengthens the possibility that $\gamma$-rays are produced in $\gamma^2$ Velorum, most likely as a result of particle acceleration in the wind collision region. The observed orbital variability is consistent with predictions from recent magnetohydrodynamic simulations, but contrasts with the orbital variability from $\eta$ Carinae, where the peak of the light curve is found at periastron.}


   \maketitle
%

\section{Introduction} \label{intro}

Particle acceleration processes in binaries are of special interest because the geometrical conditions of such systems can vary substantially along the orbit. Studying how non-thermal radiation is emitted under those different conditions provides valuable information about the mechanisms governing particle acceleration. Most of the binary systems that have been detected at high energies (HE) or very high energies (VHE) contain compact objects---the so-called $\gamma$-ray binaries, see \cite{Dubus13}---but massive binaries with strong winds have also been proposed as particle accelerators \citep{Eichler93, Dougherty00, Becker13}. Such systems are known as colliding wind binaries (CWB), composed of O or Wolf-Rayet (WR) stars. Those massive stars possess powerful, dense winds; WR stars have wind densities one order of magnitude higher than the typical O stars \citep{Crowther07}. These strong winds eventually collide, forming bow-shaped shocks in the wind collision region (WCR) delimited by two shock fronts \citep{Eichler93}. Under those conditions, diffusive shock acceleration (DSA) can occur for both protons and electrons. This may lead to the production of non-thermal radiation whose spectrum can highly depend on, among other factors, magnetic field strength and topology as well as particle density in the WCR \citep{Benaglia01, Benaglia03, Reimer06, Becker07, Reitberger14em, Grimaldo19}.

\object{$\gamma ^2$ Velorum} (or \object{WR 11}) is the closest CWB, at a distance of $d = 336^{+8}_{-7}$ pc \citep{North07} in the \object{Vela OB2} association. The spectral types of its components are WC8 and O7.5, respectively \citep{Marco00}. Its orbit is slightly elliptic with an eccentricity of $e=0.334 \pm 0.003$ \citep{North07} and a period of $P_{orb}=78.53 \pm 0.01$ days \citep{Schmutz97}.
This binary system is detected at several wavelengths. In radio, the existence of a non-thermal component is unclear \citep{Chapman99, Benaglia16}, while this system shows a strong thermal component consistent with bremsstrahlung emission \citep{Purton82, Benaglia19}. Evidence for a WCR comes from UV \citep{St.-Louis93} and X-ray observations with {\it ROSAT} \citep{Willis95}, where a wide cavity in the wind surrounding the O star is observed. The shock-cone opening angle of the wind cavity was found to be $\approx 85 ^{\circ} $ using {\it Chandra} observations \citep{Henley05}. 

Other CWBs exhibit non-thermal radio emission, summarised in a catalogue by \cite{Becker13} and references therein. Searches for $\gamma$-ray emission spatially associated with CWBs were performed in recent decades using both space instruments and ground-based telescopes \citep{Romero99, Aliu08, HESS12}. Upper limits have been obtained more recently by the {\it Fermi} Large Area Telescope ({\it Fermi}-LAT) \citep{Werner13, Pshirkov16}. In spite of the observational efforts, only $\eta$ Carinae has been firmly identified at HE \citep{Farnier11,Reitberger15, Balbo17} and probably at VHE \citep{Leser17}. $\gamma ^2$ Velorum has also been associated with a $\gamma$-ray source by \cite{Pshirkov16}, but its conclusive identification is pending since no orbital modulation has been found. Indeed, \cite{Benaglia16} find other radio sources that are positionally compatible with the excess at HE and point out that $\gamma ^2$ Velorum is the only bright radio source in the field without a negative spectral index $\alpha$ at radio frequencies (for a radio flux density defined as $S\propto \nu ^{\;\alpha}$). A more detailed description can be found in \cite{Benaglia19}. Nevertheless, \cite{Reitberger17} are able to reproduce the HE spectral energy distribution (SED) using magnetohydrodynamic (MHD) simulations, finding that emission from $\gamma ^2$ Velorum should have a hadronic origin. This HE component would arise from $\pi^0$ decay due to proton acceleration up to 1 TeV near to the WCR apex and would be orbitally modulated, thereby leading to higher fluxes during apastron passage. On the contrary, electrons would only reach 100 MeV, and therefore would not be able to produce the observed spectrum through the inverse Compton (IC) mechanism.

In this work, a search for orbital variability from $\gamma ^2$ Velorum in $\gamma$-rays was performed on $Fermi$-LAT data. Section \ref{analysis} presents the dataset and the analysis performed, while the results are presented in Section \ref{results}. Finally, discussion and a summary are included in Sections \ref{discussion} and \ref{summary}, respectively.

\section{Observations and analysis} \label{analysis}


\subsection{Analysis methods}

    The LAT is the principal $\gamma$-ray detector on the $Fermi$ Gamma-Ray Space Telescope \citep{FermiLAT}, covering the energy range between 30 MeV and more than 500 GeV. Its energy-dependent point spread function (PSF) ranges from 
$\sim$5$^{\circ}$ ($68\%$ containment) at 100 MeV to $<$0.1$^{\circ}$ above 10 GeV. In this work, observations from 2008 August 4 to 2018 November 3 are included, amounting to $\sim$10.25 years of data. The analysis was performed using {\it Fermitools-1.0.1}\footnote{This is the nomenclature for the new $Fermi$ Science Tools released through Conda. See \url{https://github.com/fermi-lat/Fermitools-conda/wiki}} on P8R3 data \citep{P8R3}. The \texttt{SOURCE} event class (evclass=128) and \texttt{FRONT}+\texttt{BACK} event type (evtype=3) were employed, together with the \texttt{P8R3$\_$SOURCE$\_$V2} instrument response functions (IRFs). All photons within a $20^\circ\times20^\circ$ region of interest (ROI) centred on $\gamma ^2$ Velorum and in the energy range $100$ MeV--$500$ GeV were selected.  Contamination from the limb of the earth was reduced by selecting events with zenith angle $<$90$^\circ$. In this work we refer to this data selection as the non-gated dataset. Two complementary data selections based on this dataset (gated and off-peak datasets) are described in Section \ref{Vela}.

We obtained the fluxes presented in this work by performing a binned maximum-likelihood fit \citep{Mattox96} via \texttt{fermipy 0.17.4}\footnote{Python package for the $Fermitools$. See \url{https://fermipy.readthedocs.io/en/latest/}} \citep{Fermipy}, with a pixel size of $0.1^\circ$ and eight energy bins per decade, and enabling energy dispersion. The source model included all sources from the 4FGL catalogue \citep{4FGL} within a $30^\circ\times30^\circ$ region centred on $\gamma ^2$ Velorum (i.e. 119 sources). The Galactic and isotropic diffuse components used are `gll$\_$iem$\_$v07.fits' and `iso$\_$P8R3$\_$SOURCE$\_$V2$\_$v1.txt', respectively. To evaluate the significance of the detection of each source, we used the following test statistic (TS):
\begin{equation}
                TS = -2\;ln\left( L_{max,0}/L_{max,1}\right)
,\end{equation}
where $L_{max,0}$ is the log-likelihood value for the model in which the source studied is removed (the null hypothesis) and $L_{max,1}$ the log-likelihood for the complete model. The larger the value of the TS is, the less likely the null hypothesis. The common threshold to claim detection of a source is TS $=25$ \citep{4FGL}.

 Using the above-described source model and data binning, a maximum-likelihood fit was performed. Firstly, the iterative \texttt{optimize} method from \textit{fermipy} was used. This method has three steps: (1) freeing the normalisation of the $N$ largest components containing a fraction of the total predicted number of counts in the model larger than \texttt{npred\_frac}. This is done according to their predicted number of events \textit{NPred} and performing a fit; (2) individually fitting the normalisation of every source not included in the first step and with \textit{NPred} > \texttt{npred\_threshold}. This is done in descending order of their \textit{NPred} values; and finally (3) individually fitting the shape of all sources with TS > \texttt{shape\_ts\_threshold} in order of their TS values. The default values used were \texttt{npred\_frac} $=0.95$, \texttt{npred\_threshold} $=1$ and \texttt{shape\_ts\_threshold} $=25$.  After this, a final fit was performed in which the normalisation was left free for all sources within $7^{\circ}$ of our target. Additionally, all parameters for sources with TS $>300$, including the spectral index for the diffuse Galactic component, were fit. The catalogue source 4FGL J0809.5$-$4714 ($5.3\sigma$ detection) is spatially compatible with the nominal position of the CWB binary, and henceforth we considered this source the $\gamma$-ray counterpart of $\gamma ^2$ Velorum, refining its position after the main fit (Section \ref{spectral}). In the particular case of 4FGL J0809.5$-$4714, its spectrum was assumed to be a simple power law (PL), and both its spectral index and normalisation parameters were freed  (the hypothesis is validated a posteriori in Section \ref{results}). This contrasts with the log-parabola description in the 4FGL catalogue of the other CWB $\eta$ Carinae. 
 
 \begin{figure*}[t]
   \resizebox{1.0\hsize}{!}
   {\includegraphics{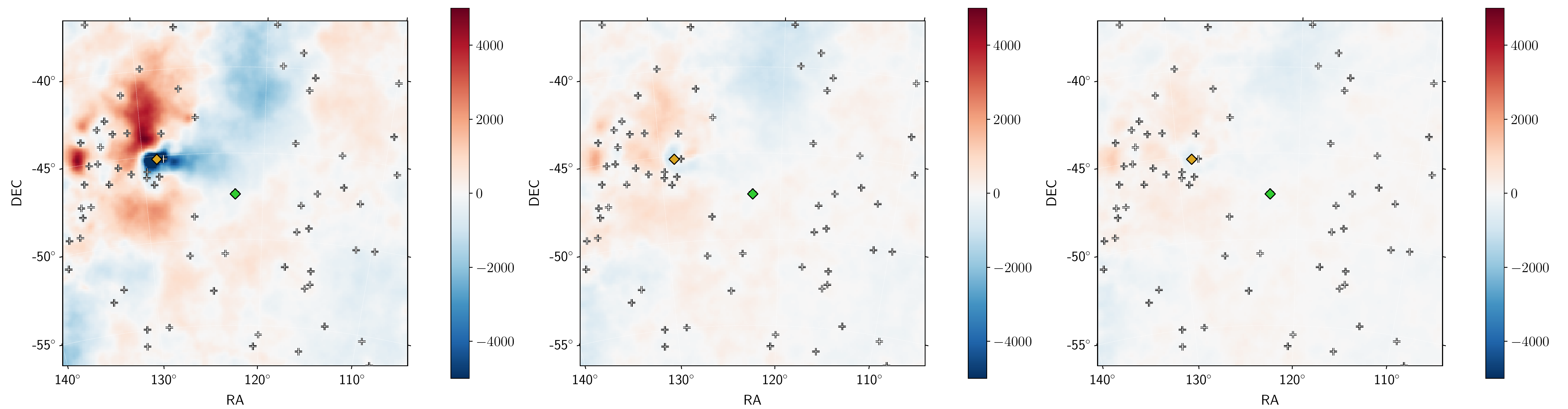}}
   \caption{Residuals of ROI in counts from the binned maximum-likelihood fit of the non-gated dataset (\textit{left}), gated dataset (\textit{centre}), and off-peak dataset (\textit{right}). Positions of all sources are represented with white crosses, while the green and yellow diamonds indicate the position of $\gamma ^2$ Velorum and the Vela pulsar, respectively. The residual for the non-gated dataset at the position of Vela is saturated for visualisation purposes of the rest of the ROI, in some some pixels reaching almost $ 5\times 10^4$ counts.}
         \label{count_residuals}
\end{figure*}
 
 \begin{figure}[t]
   \resizebox{\hsize}{!}
   {\includegraphics{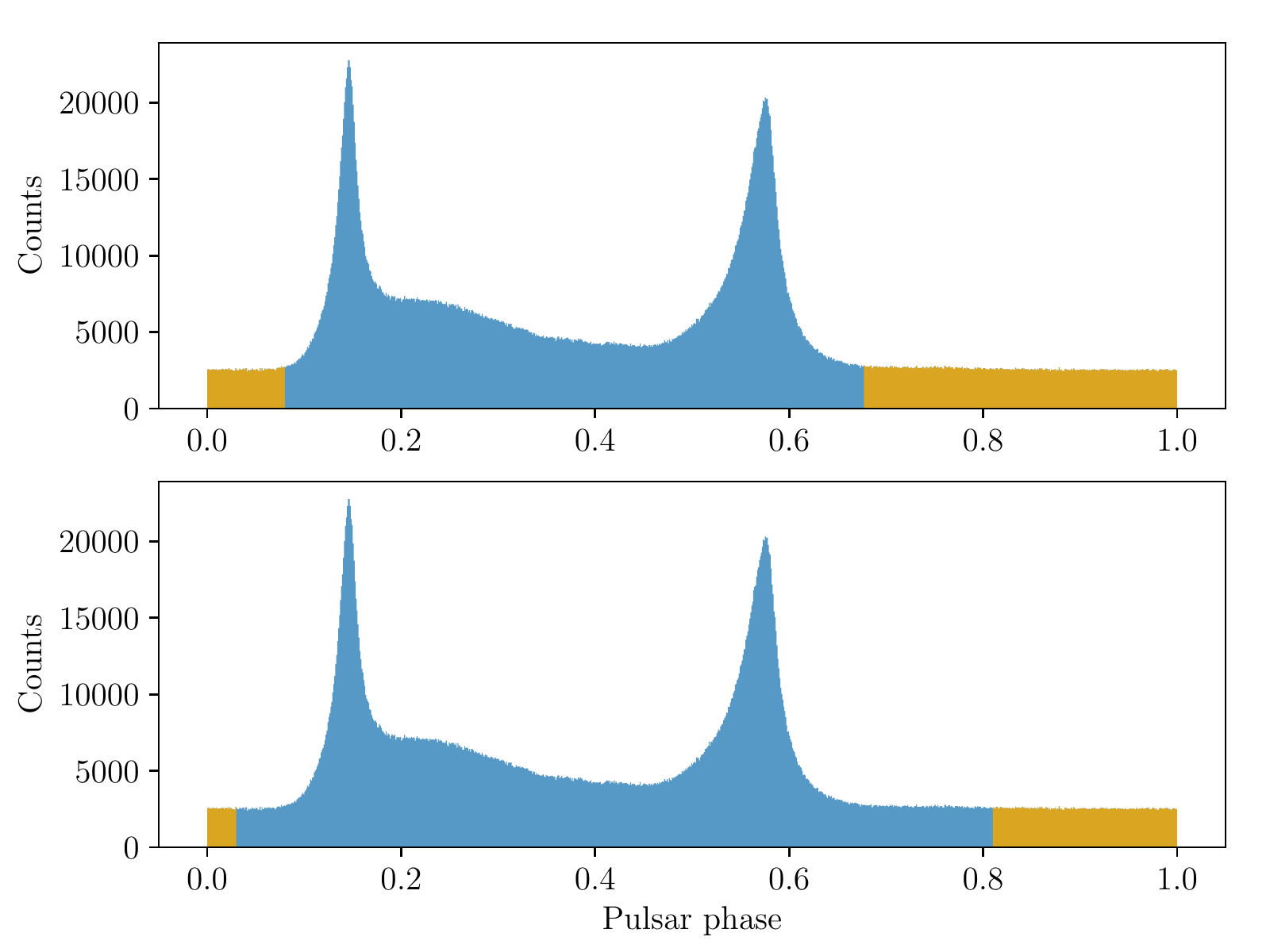}}
   \caption{Photons from within $8^{\circ}$ of $\gamma ^2$ Velorum, folded at the Vela pulsar rotational period. The gated dataset (\textit{top}) and off-peak dataset (\textit{bottom}) are shown in brown. }
              \label{pulsations}%
    \end{figure}

\subsection{Datasets} \label{Vela}

\begin{figure*}[t]
   \resizebox{1.0\hsize}{!}
   {\includegraphics{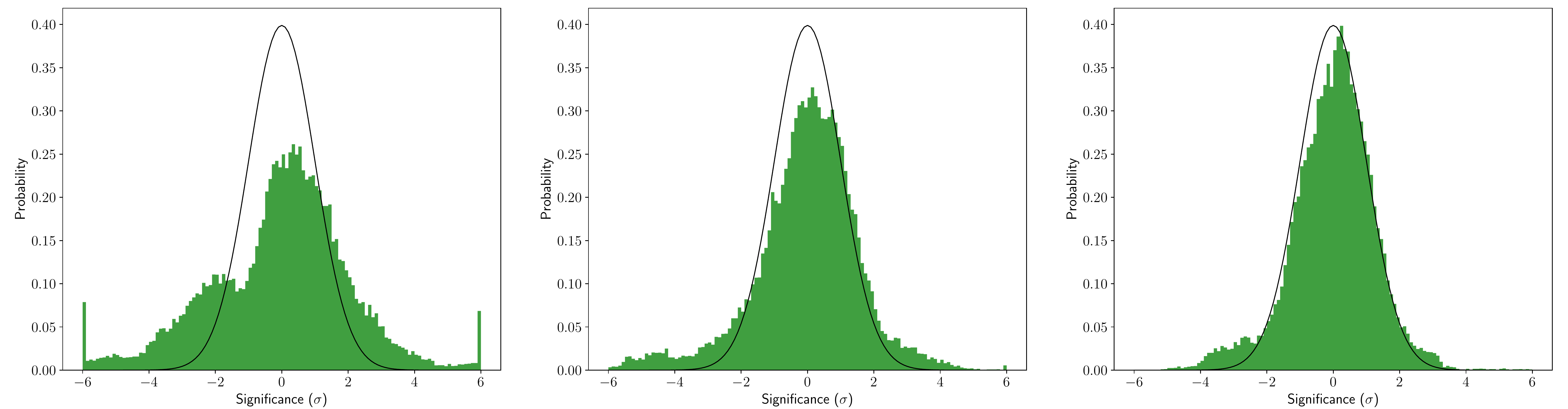}}
   \caption{Distribution of significances of the ROI residuals from the binned maximum-likelihood fit for the non-gated dataset (\textit{left}), gated dataset (\textit{centre}) and off-peak dataset (\textit{right}). A normalised Gaussian distribution is shown in black for comparison.}
         \label{sigma_distribution}
\end{figure*}

$\gamma ^2$ Velorum is separated by only $\sim$5$^{\circ}$ from the \object{Vela} pulsar, which is the brightest steady point source at HE. Since the PSF at low energies can extend up to several degrees, at those energies we expect a major contribution from Vela to the ROI. Furthermore, the usual PL with the hyper-exponential cut-off description used for pulsars is not able to account fully for its phase-averaged SED, generating extended distortions in the significance residual maps that alter their expected normal distribution (Fig. \ref{count_residuals}, left). To evaluate how the emission from Vela affects the ROI\ and whether \object{4FGL J0809.5-4714} could simply be a spurious source from incorrectly modelling the Vela pulsar, we performed a decontamination study by gating the emission from the pulsar. To achieve this, a pulse phase was assigned to each event within $8^{\circ}$ of $\gamma ^2$ Velorum with the software \texttt{TEMPO2} \citep{Hobbs06, Edwards06} and the \textit{Fermi} plug-in \citep{Ray11}. We adopted an updated ephemeris obtained using the method of \cite{Kerr15}.

Pulsations are obvious over a large region surrounding from Vela. Some level of unmodulated emission is observed in the light curve (Fig. \ref{pulsations}), which corresponds to all photons from the sources in the ROI, including diffuse emission and unmodulated emission from Vela. Such photons are uniformly distributed in the off-peak phases $\phi=[0.81-0.03]$ \citep{2PC}. Indeed, pulsations from Vela are not only found to dominate the ROI, but also within $1^{\circ}$ of the binary. Therefore a secondary analysis with a gated dataset is necessary since it can provide a reference for the fluxes obtained in the non-gated analysis to determine if emission from Vela is properly accounted for by the model (Fig. \ref{count_residuals}).

We defined an off-peak dataset including only photons between $\phi=[0.81-0.03]$ (Fig. \ref{pulsations}, bottom). Using only off-peak events substantially reduces the number of photons, which leads to lower significances and is insufficient for a phased analysis along the orbit. To eliminate the residual distortions (see Fig. \ref{sigma_distribution}) but preserve as much data as possible, we extended the off-peak selection until $2\%$ of the selected photons were above the average counts between $\phi=[0.81-0.03]$. This cleaner dataset (the gated dataset) includes pulse phases $\phi=[0.677-0.080]$ (Fig. \ref{pulsations}, top). The same study was carried out setting the limit to 1, 3, 5, and 10$\%$ to study the changes in TS detection with the exposure time and confirmed similar fluxes with the off-peak dataset, but only the $2\%$ case is presented in this work.

\section{Results} \label{results}

\subsection{Source detection} \label{spectral}

\begin{figure}
   \resizebox{\hsize}{!}
   {\includegraphics{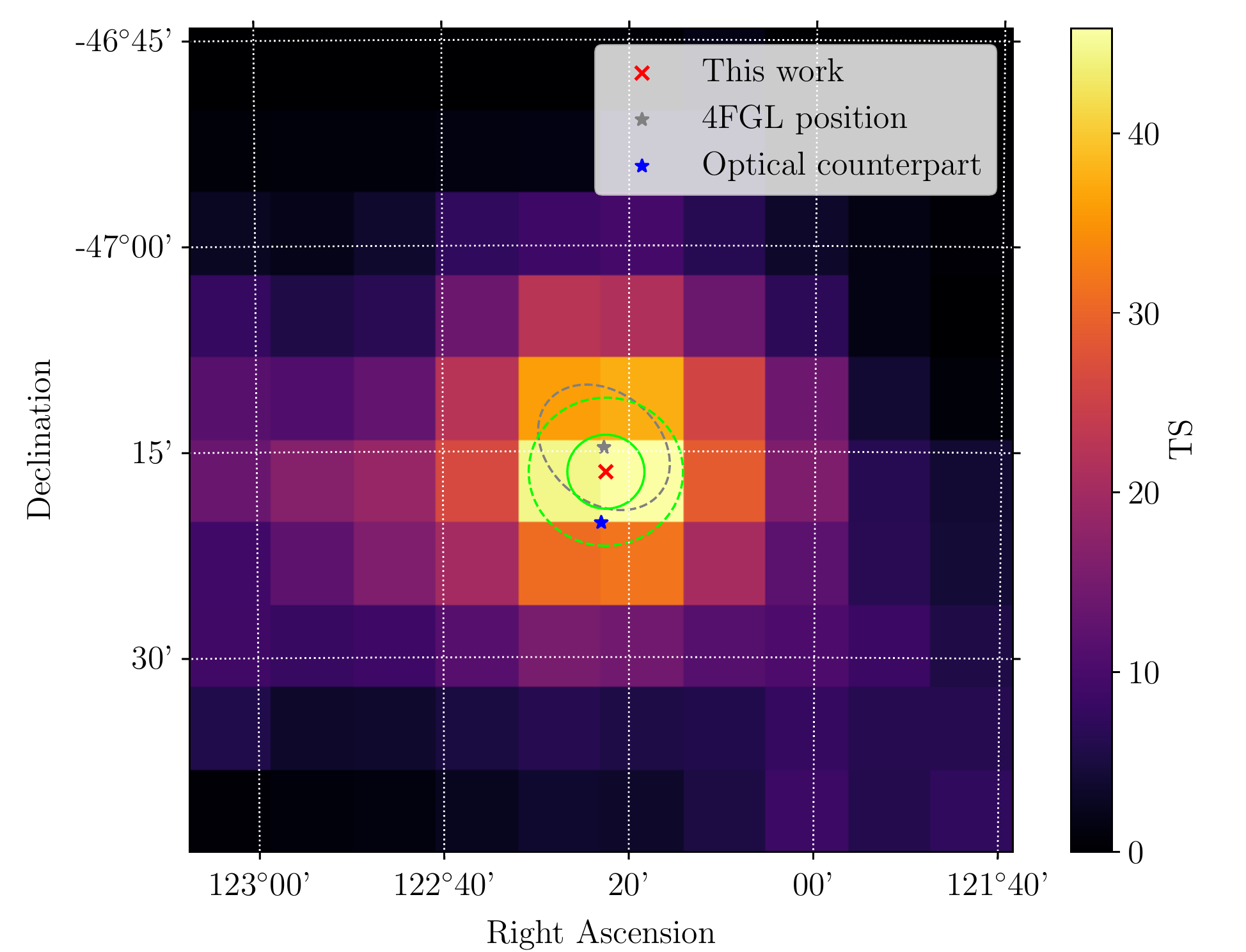}}
   \caption{Smoothed TS map with the 4FGL source and its refined position of 4FGL J0809.5-4714. The nominal position of $\gamma ^2$ Velorum is denoted with a blue star. The 68\% and 99\% containment confidence levels obtained with the \texttt{localize} algorithm are shown with solid and dashed green lines. The 4FGL $95\%$ containment error is shown in grey.}
              \label{localize}%
    \end{figure}

\begin{figure}
   \resizebox{\hsize}{!}
   {\includegraphics{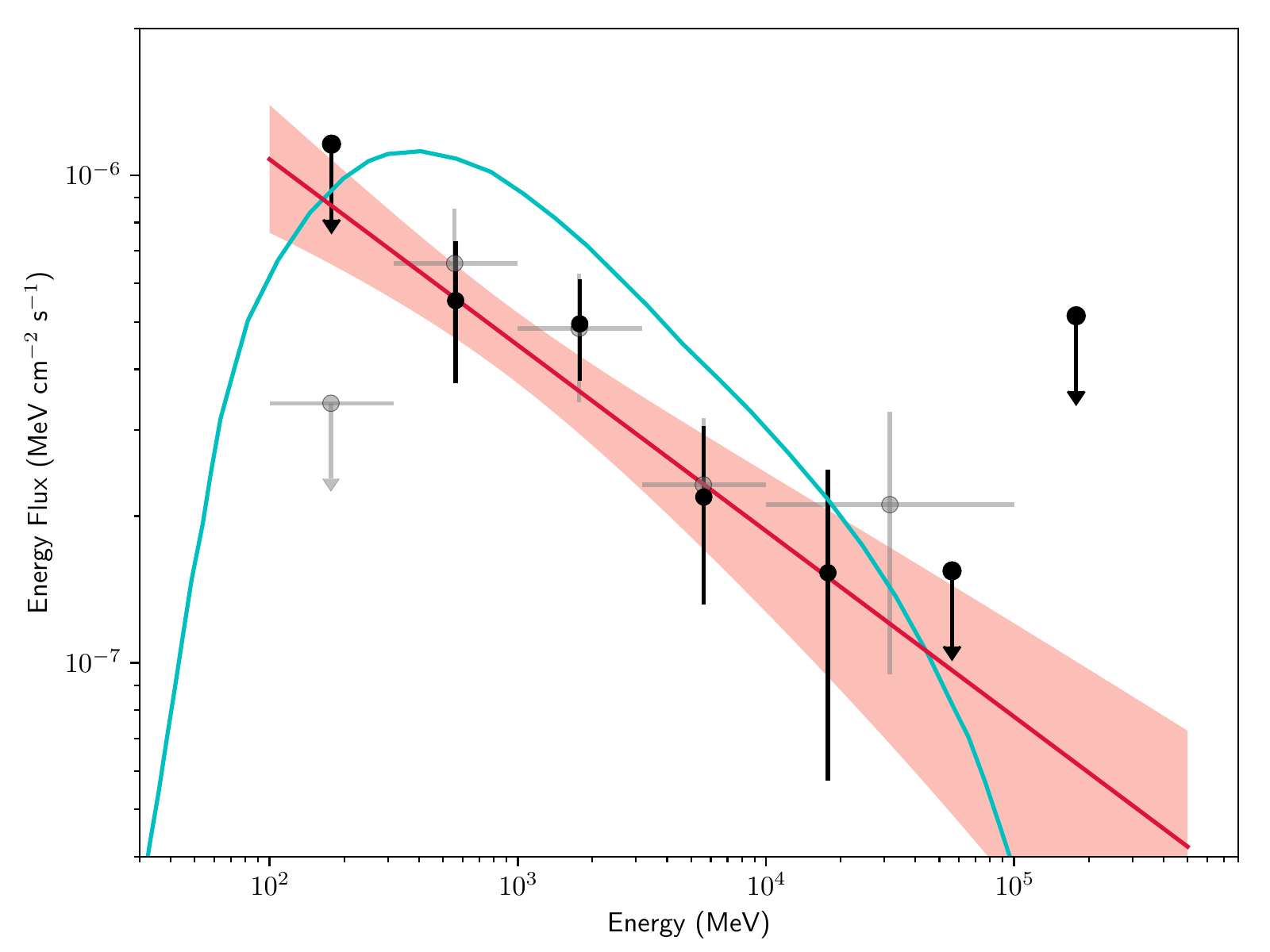}}
   \caption{Spectrum from $\gamma ^2$ Velorum. The red line corresponds to the best-fit PL spectral shape and has an uncertainty of $1\sigma$. The 95\% confidence level upper limits are used for energy bins with less than $2\sigma$ detection. Data from \citet{Pshirkov16} are shown in grey. Emission from \citet{Reitberger17} at apastron is shown in blue. The spectrum was obtained assuming $\Gamma=2.0$ per each energy bin by default with \textit{fermipy} ($\Gamma=2.39$ provides fully compatible results).}
              \label{SEDfullorbit}%
    \end{figure}

In our standard analysis with the non-gated dataset, the binary is detected  with TS$\sim$50 (Fig. \ref{localize}). The difference in the significance from previous studies or the 4FGL catalogue itself is understood to be produced by the combined effect of a larger dataset and the differences in the analysis; for example the weighted analysis performed in the 4FGL has not been reproduced. With the reduced exposure in the \textit{gated} dataset, the significance drops to TS$\sim$18. 
The fluxes from the gated, non-gated, and off-peak analyses are fully compatible with each other (Tab. \ref{values}). Besides, these fluxes are also compatible with the energy flux attributed to $\gamma ^2$ Velorum in the 4FGL catalogue ($2.58 \cdot 10^{-6}$ MeV cm$^{-2}$ s$^{-1}$). This confirms that even with a strong caveat concerning the influence of Vela in the ROI, the non-gated dataset can be safely used to estimate the source observables since the possible systematics are below the statistical noise level. The flux obtained in the maximum-likelihood analysis with the non-gated dataset is $\left( 8.07\pm 1.94 \right) \times 10^{-9}$ ph cm$^{-2}$ s$^{-1}$, which converted to energy flux is $\left(2.76\pm 0.46 \right) \times 10^{-6}$ MeV cm$^{-2}$ s$^{-1}$ integrated between 100 MeV and 500 GeV. The spectrum is modelled with a PL (see Figure \ref{SEDfullorbit}), obtaining a soft spectral index of $\Gamma = 2.39\pm0.12$. This flux is larger than in \cite{Pshirkov16}, who reported a flux between 0.1 and 100 GeV of $\left( 1.8 \pm 0.6 \right) \times 10^{-9}$ ph cm$^{-2}$ s$^{-1}$ and a PL index of $\Gamma = 2.16\pm0.20$. Our SED mainly differs from that of \citet{Pshirkov16} in a more conservative upper limit in the lower energy bin. We understand such differences in the reported flux and spectral index as arising from the PL fit, where the upper limit in the lower energy bin in \citet{Pshirkov16} made its spectral index harder while lowering the normalisation, underestimating the total flux obtained. This might be caused by the different diffuse models, source catalogues, and data used (gll\_iem\_v06.fits, 3FGL and Pass8, respectively).

The 4FGL source is only $0.09^{\circ}$ distant from the nominal position of $\gamma ^2$ Velorum. Using the tool \texttt{localize} from \textit{fermipy}, a refined TS map of the region can be generated to evaluate the position (Fig. \ref{localize}). This analysis centres the $\gamma$-ray signal at $\left( \alpha , \delta \right) = \left( 122.375^{\circ} , -47.275^{\circ} \right)$, $0.062^{\circ}$ from the binary, with an error region of radius $0.046^{\circ}$, $0.074^{\circ}$, and $0.092^{\circ}$ for $68\%$, $95\%$, and $99\%$ confidence region. Consequently the tension between the optical counterpart and the HE source position is reduced. The refined location is used in subsequent analysis.

\subsection{Orbital variability analysis} \label{orbit}

For an unambiguous identification of the $\gamma$-ray source with $\gamma ^2$ Velorum, finding phase-locked orbital variability is crucial. Therefore, we assigned to each photon a phase according to the binary orbital period of $78.53$ days. As a reference for the periastron passage ($\phi=0.0$), we define $T_0=50120.9$ MJD as in \citet{North07}. The non-gated dataset can be split into two broad orbital phases: periastron ($\phi=0.0-0.25\;\&\;\phi=0.75-1.0$) and apastron ($\phi=0.25-0.75$). The analysis described in Section \ref{analysis} is applied to both orbital phases (Tab. \ref{values}). Emission is found mainly during apastron, with an energy flux of $\left( 3.57 \pm 0.66 \right) \times 10^{-6}$ MeV cm$^{-2}$ s$^{-1}$ and TS$\sim 41$, while in periastron the flux is reduced to $\left( 1.56 \pm 0.58\right) \times 10^{-6}$ MeV cm$^{-2}$ s$^{-1}$, with a detection significance of TS $\sim11$ (Tab. \ref{values}). No significant variation of the PL index over the orbit is detected. To evaluate the flux variability, we used the same method introduced in the 2FGL for $\gamma$-ray sources \citep{2FGL}. In this work we adapt this method to a phase-binned analysis,

\begin{equation}
                TS_{var}=2\sum_i^{N_{bins}} \left[ \log \mathcal{L}_i\left(F_i\right) - \log \mathcal{L}_i\left(F_{const}\right) \right]
        \label{eq:tsvar}
        ,\end{equation}

where we use the model from the full-orbit analysis for consistency. The $TS_{var}$ statistic follows a $\chi ^2$ distribution with the appropriate number of degrees of freedom (d.o.f.). For two bins (1 d.o.f.) we obtain $TS_{var}=5.91$, which corresponds to a p-value of 0.015 ($\sim 2.4\sigma$).

Since the periastron/apastron flux change is marginally significant, we cannot conclusively establish the variability of HE emission with the orbit of the binary. However, the results show that there are hints of a physical relation between the $\gamma$-ray source 4FGL J0809.5$-$4714 and $\gamma ^2$ Velorum. To check the reliability of this result, we performed the same analysis on the gated dataset. Even if no influence from the Vela pulsar emission on the variability itself is expected\footnote{For caveats on \textit{Fermi}-LAT temporal analyses see \url{https://fermi.gsfc.nasa.gov/ssc/data/analysis/LAT_caveats_temporal.html}} ($P_{Vela\_rot}=89.4$ ms $<<P_{orb}$, \citep{2PC}) and if the low statistics provide larger errors, the gated dataset is cleaner. The result obtained shows the same trend; there is a larger flux at apastron than in periastron and compatible values with the non-gated analysis (Tab. \ref{values}). 

\begin{table*}
\caption{Spectral results for 4FGL J0809.5-4714.}             
\label{table:1}      
\centering                          
\begin{tabular}{c c c c}        
\hline\hline   
Orbital phase & TS & Energy flux ($10^{-6}$ MeV cm$^{-2}$ s$^{-1}$) & $\Gamma$ \\  [0.5ex]
\hline\hline                   
   Non-gated dataset & $49.96$ & $2.76 \pm 0.46$ & $2.39 \pm 0.12$ \\      
   Periastron & $10.59$ & $1.56 \pm 0.58$    & $2.30 \pm 0.22$ \\
   Apastron & $40.85$ & $3.57 \pm 0.66$     &  $2.35 \pm 0.13$ \\
\hline      
        Gated dataset & $18.29$ & $2.63 \pm 0.69$ & $2.54 \pm 0.17$ \\                            
        Periastron (GD) & $2.87$ & $1.45 \pm 0.96$    & $2.54$ (fixed)  \\
        Apastron (GD) & $21.88$ & $4.03 \pm 0.97$     &  $2.54$ (fixed) \\
\hline  
        Off-peak dataset & $10.80$ & $2.55 \pm 0.90$ & $2.42 \pm 0.26$ \\
\hline  
\end{tabular}
\label{values}
\end{table*}

Additionally, we studied the photon flux along the orbit using the non-gated dataset. As a consequence of the low significance per bin, the results obtained by dividing the orbit into smaller bins might provide limited information. However, the higher time resolution could provide hints about how the photons are emitted along the orbit, for example if they cluster around a specific orbital phase. For this purpose, we divided the orbit into 4 ($\Delta \phi$ $=0.25$), 6 ($\Delta \phi$ $=0.167$), and 8 ($\Delta \phi$ $=0.125$) bins. For each bin, we corrected the fluxes by a factor $1/N_{bins}$ to account for the reduced exposure relative to the whole dataset. The average variations of the exposure from bin to bin are found to be $\sim 2 \%$. We studied the impact of such changes to the present results, without any significant deviation of the fluxes reported. 

The light curves were studied with two different approaches. First, an analysis was performed in each bin using the normalisation and spectral indexes obtained for each source in the ROI analysis performed on the non-gated dataset. Only the normalisation parameter for $\gamma^2$ Velorum is left free, with $\Gamma$ constant. However, this approach can be very sensitive to statistical fluctuations from Vela. 

To account for any possible systematics, we also performed a secondary analysis with the normalisations of the Vela pulsar, the Galactic, and the isotropic diffuse components as free parameters. Even if we do not expect any variations from other sources at the timescales of the orbit, the inhomogeneities in the ROI induced by the cuts in the different bins might influence those sources which contribute to the whole ROI. Such a conservative approach has been used in the \textit{Fermi}-LAT source catalogues since \citet{2FGL}.

\begin{figure*}
   \resizebox{\hsize}{!}
   {\includegraphics{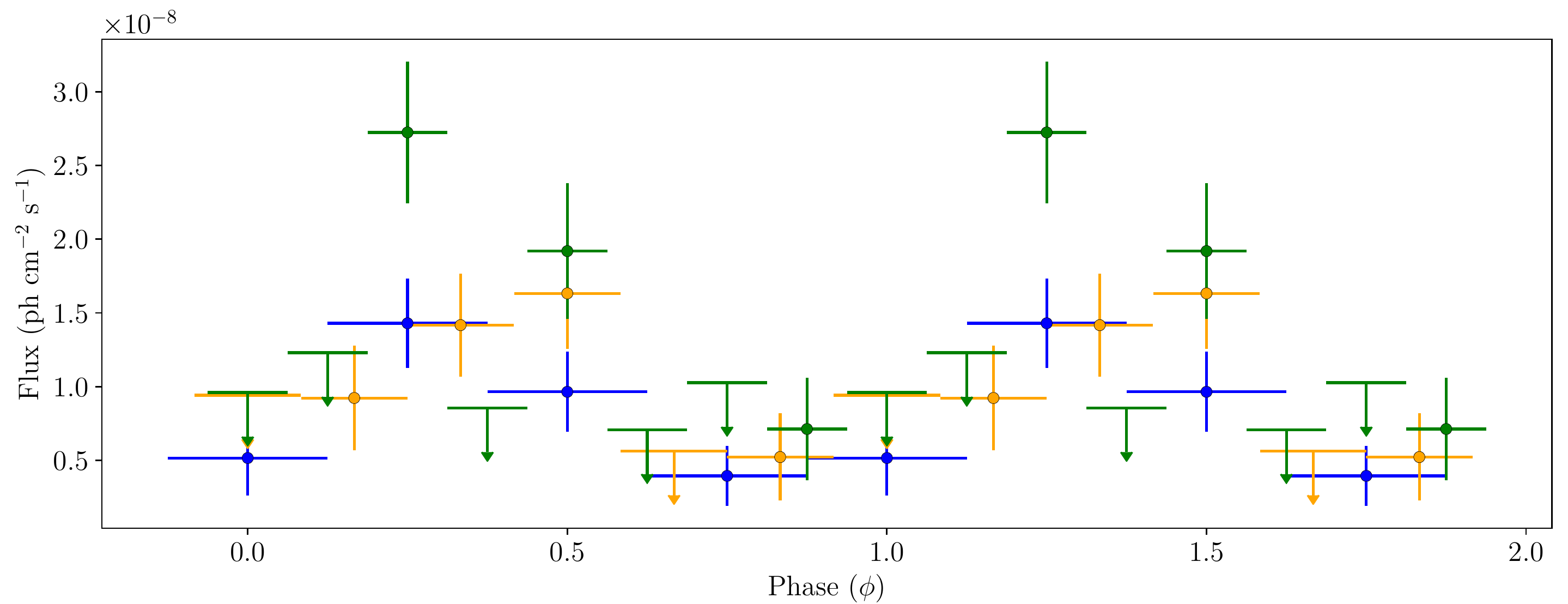}}
   \caption{Light curve for 4FGL J0809.5$-$4714 when folded on the orbital period of $\gamma ^2$ Velorum. Orbits are divided in 4 (blue), 6 (orange), and 8 (green) bins. Periastron occurs at phase $\phi = 0$. Data are duplicated to show two orbits. The 95\% confidence level upper limits are used for bins with less than $2\sigma$ detection.}
              \label{lightcurve}%
    \end{figure*}

The results for the first case are shown in Figure \ref{lightcurve}, where all light curves peak slightly before apastron (in the phase range $\phi = \left[ 0.25 - 0.5\right]$). This behaviour agrees with the previous result. To evaluate such variability, we used the same $TS_{var} $ method introduced in Equation \ref{eq:tsvar}. We obtain probabilities for the non-variability hypothesis of $2.23 \cdot 10^{-2}$ , $3.14 \cdot 10^{-3}$, and $1.34 \cdot 10^{-5}$ for the 4, 6, and 8 bins, respectively (see Table \ref{tab:variability}). Similar results are found using a $\chi^2$ test. 
In those cases the probabilities for the non-variability hypothesis are $2.35 \cdot 10^{-2}$ , $1.80 \cdot 10^{-3}$, and $3.73 \cdot 10^{-5}$, respectively.  Going to shorter bins (e.g. 10, 12) reverses the trend of increasing significance as the signal becomes too diluted for detection in single bins.

\begin{table*}
\caption{Variability studies using the $TS_{var}$ and $\chi^2$ methods for the non-conservative analysis (i.e. freeing only the normalisation parameter for $\gamma^2$ Velorum) and the conservative analysis (i.e. also freeing the normalisation parameters for Vela and the diffuse components). }             
\label{table:2}      
\centering    
\begin{tabular}{c |c c | c c | c c | c c |}        
\cline{2-9}
 & \multicolumn{4}{c}{Non-conservative} & \multicolumn{4}{|c|}{Conservative} \\ [0.5ex]
\cline{2-9}
 &  $TS_{var}$ & p-value& $\chi^2$& p-value & $TS_{var}$ & p-value& $\chi^2$ & p-value\\  [0.5ex]
\hline\hline  
    4 bins (3 d.o.f.)& $9.60$ & $2.23 \cdot 10^{-2}$ & $9.48$ & $2.35\cdot 10^{-2}$& $2.19$ & $0.53$ & $2.13$ & $0.50$\\      
    6 bins (5 d.o.f.)& $17.85$ & $3.14 \cdot 10^{-3}$    & $19.15$ &$1.80 \cdot 10^{-3}$& $8.05$ & $0.15$ & $9.70$ & $0.08$\\
    8 bins (7 d.o.f.)& $34.58$ & $1.34 \cdot 10^{-5}$     &  $32.20$& $3.73 \cdot 10^{-5}$& $12.81$ & $0.08$ & $15.47$ & $0.03$\\

\hline  
\end{tabular}
\label{tab:variability}
\end{table*}

The second analysis scheme is more conservative but could mask real variability of the binary by mixing in statistical fluctuations from the much brighter Vela pulsar and diffuse components. The results for the $TS_{var}$ and $\chi^2$ statistics can be found in Table \ref{tab:variability}. In this conservative analysis the variability is no longer significant, but the larger fluxes are still found around apastron.

\section{Discussion} \label{discussion}

\subsection{$\gamma ^2$ Velorum as a particle accelerator} \label{acceleration}

The results obtained provide hints of variability. Therefore a physical relation between this particular binary and the HE signal is supported, potentially discarding other nearby non-thermal radio sources as the origin of the $\gamma$-rays \citep{Benaglia16, Benaglia19}. The system is studied and modelled by \citet{Reitberger17} using 3D MHD simulations. Although these authors do not use the exact values of the spectrum from \citet{Pshirkov16} owing to an erratum (see the deviation of the model with respect the data in Figure \ref{SEDfullorbit}), they report similar flux levels. The spectrum from $\gamma ^2$ Velorum could be explained as emission arising only from the apastron passage, but also as the average flux along the orbit (with lower emission at intermediate phases, for example $\phi = 0.3$). Periastron emission could not explain the observed flux in any case, even after tuning the free parameters of injection rate or the diffusion coefficient. Therefore at $\phi = 0$ the flux is predicted to be significantly lower than at other orbital phases. This can be explained because towards periastron, $\gamma$-ray fluxes would be reduced resulting from (1) a smaller volume of the WCR, (2) protons reaching the WCR with less velocity, and (3) increasing Coulomb losses because of higher densities. The larger emission at apastron and smaller fluxes at periastron indicated in our results are consistent with this particular feature of the predicted signal. Thus, the flux variability observed along the orbit can constrain the parameters for the proton injection rate $\eta$ and the normalisation of the diffusion coefficient $D_0$. Unfortunately, no extensive comparison can be performed because of a lack of quantitative information. 

Since in their simulations electrons could not exceed energies of 100 MeV in the outer wings of the WCR, \citet{Reitberger17} also claim that the emission is hadronic in origin (i.e. photon emission produced by $\pi ^0$-decay). This result would agree with previous studies pointing to hadronic emission also in close (i.e. short stellar distances) CWBs \citep{Reitberger14em}. Distinguishing between non-thermal radiation production mechanisms is a non-trivial issue. Photons have no signature from the radiative mechanism and neutrino fluxes expected from CWBs are too low to be unambiguously detected with the present generation of neutrino observatories, even for \object{$\eta$ Carinae} \citep{Gupta17}. Thus, we rely on spectral models. Cut-offs may be characteristic of SEDs produced via hadronic processes, although they are not exclusive. A sharp cut-off is expected below 100 MeV and another is expected at $\sim$100 GeV, which might be caused by a maximum energy of $\sim$1 TeV of the protons in this particular system. This maximum energy of the protons is not significantly affected by  wind advection. At such energies, protons were found to be highly dominated by diffusion \citep{Reitberger17}. Unfortunately, the low flux expectations and the sensitivity of \textit{Fermi}-LAT in those regimes prevent us from confirming such spectral features (Fig. \ref{SEDfullorbit}) or a possible effect of photon-photon absorption above $50$ GeV. Above 100 GeV we can only constrain the flux for $\gamma ^2$ Velorum with upper limits.

Future missions studying the sky at HE may be able to confirm the hints of variability reported in this work, while new observatories such as the Cherenkov Telescope Array (CTA) might be able to shed light on the extent of the component to VHEs \citep{Paredes13, Chernyakova19, CTA19}.

\subsection{Comparison with $\eta$ Carinae} \label{etacar}
If the source 4FGL J0809.5$-$4714 can  be identified as $\gamma^2$ Velorum, it would be the second CWB showing $\gamma$-ray emission. However, the observed behaviour in Section \ref{orbit} is in stark contrast to $\eta$ Carinae.  While this work finds evidence for a flux maximum at apastron (see Tab. \ref{values}), $\eta$ Carinae peaks during the periastron passage \citep{Reitberger15, Balbo17}. Additionally, emission from $\gamma^2$ Velorum can be explained with a single component, while the non-thermal radiation of $\eta$ Carinae  is generally accepted to arise from different components: a leptonic signal from IC below 10 GeV and a hadronic component from $\pi ^0$ decay above 10 GeV \citep{Farnier11}. While the IC emission is less variable, this hadronic component varies in flux by half an order of magnitude along the orbit.  The potential flux variability in $\gamma ^2$ Velorum is of a similar scale (Fig. \ref{lightcurve}).

The different properties of both systems can be connected to conditions in the WCR. According to \citet{Reitberger17}, electron acceleration in $\gamma ^2$ Velorum only reaches 10 MeV at the apex and 100 MeV in the outer wings of the WCR as a consquence of the strong radiation field. Therefore, while $\pi ^0$ decay can be responsible for the observed spectrum (see Section \ref{acceleration}), IC losses in the WCR prevent electrons from reaching higher energies in this binary and its leptonic component lies out of the $Fermi$-LAT detection capabilities. The situation is completely different for $\eta$ Carinae, where different models allow both electron and proton populations to reach HE along its orbital period of $\sim$ 5.5 years. Whether protons can be accelerated up to PeV energies remains an open question \citep{Gupta17}. The geometry of the systems was expected to be one of the most relevant factors in the emission patterns of CWB, thus in previous studies HE emission was predicted from long period binaries. Most CWBs show non-thermal radio emission and these long-orbit systems would be wide enough to avoid radiative braking of the accelerated electrons in the WCR \citep{Reimer06, Reitberger14em}. However, the $\gamma$-ray variability in $\gamma ^2$ Velorum could confirm that short period systems can also accelerate particles to HE.

Surprisingly, the lack of synchrotron radiation is a common signature of both CWBs, in comparison with the absence of $\gamma$-ray emission from studied systems with non-thermal radio emission \citep{Werner13,Pshirkov16}. This lack of non-thermal radio signatures is presumably due to free-free absorption \citep{Benaglia19}, although synchrotron self-absorption might also be relevant in short period systems such as $\gamma ^2$ Velorum \citep{Becker18}. Traditionally it was expected that CWBs would have HE radiation through the IC emission mechanism  \citep{Benaglia03, Reimer06}, while more recently numerical simulations pointed to hadronic emission under certain conditions in systems with short stellar separations \citep{Reitberger14em}. Therefore, binaries without clear non-thermal radio emission were not considered as particle accelerating systems \citep{Becker13}, but the particular case of HE from $\gamma ^2$ Velorum supports the idea that systems without detected non-thermal radio emission can also be particle accelerators, as detailed in \citet{Reitberger17} and the present work.

\section{Summary} \label{summary}
The $\gamma$-ray source reported by \cite{Pshirkov16} and associated with $\gamma^2$ Velorum is present in the 4FGL catalog. To identify this source as $\gamma^2$ Velorum, we used more than ten years of \textit{Fermi}-LAT data to search for orbital modulation of the signal. Additionally, we studied the influence of emission from the Vela pulsar since its emission is predominant in the ROI, strongly challenging any analysis of faint sources. However, results obtained gating pulsar emission are compatible with the analysis of the non-gated dataset.

We find hints of orbital variability, with larger fluxes at apastron ($\phi = 0.5$). The values of energy flux between periastron and apastron half-orbits differ at $2.4\sigma$ confidence level. Besides, a light curve with shorter orbital bins shows a similar behaviour. We employed a likelihood test to evaluate the significance of the variability, which provides a probability for the non-variability hypothesis in an eight-bin, phase-folded light curve of $0.08$ and $1.34\cdot10^{-5}$ for a conservative and non-conservative analysis, respectively. The results support the models from \citet{Reitberger17}, who predicted high/low flux states at apastron/periastron using MHD simulations. As a result, a hadronic model remains the preferred explanation for the $\gamma$-ray emission of the binary. 

The evidence of orbital variability, together with the spatial coincidence of the signal with $\gamma^2$ Velorum makes this binary the most likely counterpart. Therefore, this would be the second CWB detected in the GeV regime after $\eta$ Carinae.

\begin{acknowledgements}
The Fermi LAT Collaboration acknowledges generous ongoing support from a number of agencies and institutes that have supported both the development and the operation of the LAT as well as scientific data analysis. These include the National Aeronautics and Space Administration and the Department of Energy in the United States, the Commissariat \`{a} l'Energie Atomique and the Centre National de la Recherche Scientifique / Institut National de Physique Nucl\'{e}aire et de Physique des Particules in France, the Agenzia Spaziale Italiana and the Istituto Nazionale di Fisica Nucleare in Italy, the Ministry of Education, Culture, Sports, Science and Technology (MEXT), High Energy Accelerator Research Organization (KEK) and Japan Aerospace Exploration Agency (JAXA) in Japan, and the K. A. Wallenberg Foundation, the Swedish Research Council and the Swedish National Space Board in Sweden. Additional support for science analysis during the operations phase from the following agencies is also gratefully acknowledged: the Istituto Nazionale di Astrofisica in Italy and the Centre National d'Etudes Spatiales in France. This work was performed in part under DOE Contract DE-AC02-76SF00515. The authors would like to thank M. Kerr for providing the updated ephemeris for the Vela pulsar.
\end{acknowledgements}

%
%

\bibliographystyle{aa}
\bibliography{sample.bib}

\end{document}